\documentclass[tradiabstract]{aa}

\usepackage{graphicx}
\usepackage{longtable}
\usepackage{pdflscape}

\usepackage{txfonts}

\newcommand{\au}{{\sc au}}
\newcommand{\micron}{$\mu$m}

\renewcommand{\ion}[2]{#1\,{\sc #2}}
\newcommand{\etal}{et al.}
\newcommand{\dd}{\mathrm{d}}

\newcommand{\Mearth}{M$_\oplus$}
\newcommand{\Rearth}{R$_\oplus$}
\newcommand{\Lya}{Ly$\alpha$}
\newcommand{\hst}{\emph{HST}}
\newcommand{\iue}{\emph{IUE}}

\newcommand{\fluxunitint}{erg~s$^{-1}$~cm$^{-2}$}
\newcommand{\G}{\mathrm{G}}
\newcommand{\kB}{\mathrm{k_B}}

\titlerunning{\hst/STIS \Lya\ observations of GJ~436}
\authorrunning{Ehrenreich, Lecavelier des Etangs \& Delfosse}

\begin{document}

	\title{\hst/STIS Lyman-$\alpha$ observations of the quiet M dwarf \object{GJ~436}}
	\subtitle{Predictions for the exospheric transit signature of the hot neptune \object{GJ~436b}\thanks{Based on observations made with the Space Telescope Imaging Spectrograph on board the \emph{Hubble Space Telescope} (Cycle~17 program GO/DD 11817)}}

	\author{D.~Ehrenreich\inst{1}, A.~Lecavelier des Etangs\inst{2} \& X. Delfosse\inst{1}} 

	\offprints{D.~Ehrenreich}

    \institute{
    	Institut de plan\'etologie et d'astrophysique de Grenoble (IPAG), Universit\'e Joseph Fourier-Grenoble~1, CNRS (UMR~5274), BP~53 38041 Grenoble CEDEX~9, France, \email{david.ehrenreich@obs.ujf-grenoble.fr}
	\and
	Institut d'astrophysique de Paris, Universit\'e Pierre et Marie Curie, CNRS (UMR~7095), 98\,bis boulevard Arago, 75014 Paris, France}

    \date{}

    \abstract{Lyman-$\alpha$ (\Lya) emission of neutral hydrogen ($\lambda1\,215.67$~\AA) is the main contributor to the ultraviolet flux of low-mass stars such as M dwarfs. It is also the main light source used in studies of the evaporating upper atmospheres of transiting extrasolar planets with ultraviolet transmission spectroscopy. However, there are very few observations of the \Lya\ emissions of quiet M dwarfs, and none exist for those hosting exoplanets. Here, we present \Lya\ observations of the hot-neptune host star GJ~436 with the \emph{Hubble Space Telescope} Imaging Spectrograph (\hst/STIS). We detect bright emission in the first resolved and high quality spectrum of a quiet M dwarf at \Lya. Using an energy diagram for exoplanets and an $N$-body particle simulation, this detection enables the possible exospheric signature of the hot neptune to be estimated as a $\sim11\%$ absorption in the \Lya\ stellar emission, for a typical mass-loss rate of $10^{10}$~g~s$^{-1}$. The atmosphere of the planet GJ~436b is found to be stable to evaporation, and should be readily observable with \hst. We also derive a correlation between X-ray and \Lya\ emissions for M dwarfs. This correlation will be useful for predicting the evaporation signatures of planets transiting other quiet M dwarfs.}
    

    \maketitle

\section{Introduction}
%=====================
\label{sec:intro}

Planetary transits provide golden opportunities to probe the atmospheres of extrasolar planets. The atmospheric signals detected by this method in the visible (e.g., Charbonneau \etal\ 2002) are usually tenuous, while the infrared detections of atmospheric signatures remain disputed (e.g., Gibson, Pont \& Aigrain 2010). In the ultraviolet (UV), planetary transits of the hot giant planet \object{HD~209458b} are detected from conspicuous spectroscopic signatures, which are seen as absorption in the emission lines originating from the stellar chromosphere and transition region between the chromosphere and the corona (see Ehrenreich 2010 for a review). The strongest signature is a $(15\pm4)\%$ absorption in the resolved stellar Lyman-$\alpha$ (\Lya) emission line of neutral atomic hydrogen (\ion{H}{i}) detected in medium-resolution spectra taken with the Space Telescope Imaging Spectrograph (STIS) on the \emph{Hubble Space Telescope} (\hst) (Vidal-Madjar \etal\ 2003; see also Ehrenreich \etal\ 2008). Linsky \etal\ (2010) present new observations with the Cosmic Origins Spectrograph (COS) on \hst. They report a $(7.8\pm1.3)\%$ absorption in the \ion{C}{ii} line, confirming the previous result of Vidal-Madjar \etal\ (2004).

These distinctive UV signatures, compared to those for the visible transit of the whole planet ($1.6\%$; Charbonneau \etal\ 2000; Henry \etal\ 2000), require the presence of an extended \ion{H}{i} upper atmosphere, or exosphere, to the planet. This envelope must be evaporating because (i) it must fill the planetary Roche lobe to account for the size of the absorption, and (ii) the \ion{H}{i} atoms must be accelerated by the stellar radiation pressure beyond the escape velocity of the planet to account for the absorption in the wings of the \Lya\ line (Vidal-Madjar \etal\ 2003). The presence of elements heavier than hydrogen (C, O, Si) at high altitudes requires an hydrodynamic escape (or blow-off) of the upper atmosphere, with the flow of escaping \ion{H}{i} carrying heavier species up to the Roche limit (Vidal-Madjar \etal\ 2004; Linsky \etal\ 2010), where they are swept away by radiation pressure (Lecavelier des Etangs, Vidal-Madjar \& D\'esert 2008).

New cases of atmospheric escape have been recently reported for two hot jupiters. Lecavelier des Etangs \etal\ (2010) used \hst/ACS to detect the \ion{H}{i} exosphere of \object{HD~189733b} at \Lya. Fossati \etal\ (2010) observed the highly irradiated planet host star \object{WASP-12} with \hst/COS. They interpret the tentative detection of extra absorption in the metal lines of the star WASP-12 during the transit as signs of mass loss. Thus, atmospheric evaporation could be a common phenonemon among close-in planets.

These results have inspired a comprehensive modelling effort (e.g., Garc\'\i a-Mu\~noz 2007) suggesting that the upper atmospheres of close-in giant planets must be heated by the stellar X-rays, and both extreme and far UV (Cecchi-Pestellini \etal\ 2009) to about 10\,000~K. This temperature has been inferred in the thermosphere of HD~209458b from the detection of excited hydrogen atoms (Ballester, Sing \& Herbert 2007). The typical atmospheric mass-loss rate $\dot{m}$ derived from models is between $10^{10}$ and $10^{11}$~g~s$^{-1}$. In this case, evaporation should not severely impact the stability of giant planets' atmospheres. On the other hand, it might be more important for lower-mass planets such as hot neptunes and super-earths, if their mass-loss rates are of the same order as for more massive planets.

In this article, we evaluate whether mass loss from a hot neptune such as GJ~436b could be detected with current instrumentation, and what its UV transit signature could be. GJ~436b is the first transiting Neptune-mass planet discovered (Butler et al. 2004). It orbits a close-by (10.2~pc) M2.5{\sc v} dwarf with $V=10.68$, and triggered unprecedented interest when Gillon \etal\ (2007a) detected a photometric transit of the planet, inferring a true mass of $23.17$~\Mearth and a radius of $4.22$~\Rearth, which are similar to those of either Uranus or Neptune. (The planet properties, taken from Torres (2007), are summarized in Table~1 along with some stellar parameters.) However, GJ~436b differs significantly from the Solar System ice giants because of its moderately eccentric orbit with a semi-major axis of 0.029 astronomical unit (\au). At this distance from its $0.026$-L$_\odot$ parent star, the planet indeed receives $\sim 30\,000$ times more flux than Neptune receives from the Sun, and has a blackbody brightness temperature of $T_B=717\pm35$~K (Demory \etal\ 2007). Using the \emph{Spitzer Space Telescope}, Gillon et al. (2007b) inferred a planet mean density of $\sim 1.7$~g~cm$^{-3}$, too weak to be that of a body composed exclusively of water ice, so that GJ~436b must possess a hydrogen/helium envelope that would account for $\sim 10\%$ to 20\% of the planet's mass (Fortney et al. 2007; Figueira \etal\ 2009). The detection of this envelope with transmission spectroscopy could validate the internal structure models predicting its presence. 

To assess this possiblity, it is necessary to derived both the stellar \Lya\ emission line profile and brightness. We present in Sect.~\ref{sec:obs} the observations that allowed us to calculate the stellar emission as seen from Earth. We had to correct for the interstellar medium (ISM) absorption before estimating the intrinsic stellar \Lya\ brightness: this analysis is presented in Sect.~\ref{sec:flux}. In Sect.~\ref{sec:rate}, we estimate a range of plausible mass-loss rates in GJ~436b's upper atmosphere. This range serves as an input to an atmospheric escape model used to estimate the transit signature of the evaporating hydrogen envelope of the planet in the UV (Sect.~\ref{sec:model}). To our knowledge, these observations are the first resolved accurate  measurement of a quiet M dwarf's \Lya\ emission. Hence, we discuss in Sect.~\ref{sec:Mdwarfs} how they compare to archival \Lya\ observations of active M dwarfs.

\begin{table}
%************
\caption{\label{tab:gj436b} Properties of GJ~436 and its planet}
\begin{tabular}{lrc}
\hline\hline  
Parameter                                               & Value                  & Ref. \\
\hline
\multicolumn{3}{c}{The planet} \\
\hline
Mass $M_p$ (\Mearth)                             & $23.17\pm0.79$         & (1) \\
Radius $R_p$ (\Rearth)                           & $4.22^{+0.09}_{-0.10}$ & (1) \\
Mean density $\rho_p$ (g~cm$^{-3}$)              & $1.69^{+0.14}_{-0.12}$ & (1) \\
Gravity $g_p$ (m~s$^{-2}$)              & $12.8\pm0.8$           & (1) \\
Semi-major axis $a_p$ (\au)                      & $0.02872\pm0.00027$    & (1) \\
Brightness temperature $T_B$ (K)                 & $717\pm35$             & (3) \\
\hline
\multicolumn{3}{c}{The star} \\
\hline
Distance $d_\star$ (pc)                 & $10.23\pm0.24$         & (4) \\
X/EUV luminosity                        &                        &     \\
$\quad\log (L_\mathrm{X/EUV}~[\mathrm{erg~s^{-1}}]) $    & $26.85^{+0.65}_{-0.89}$& (2) \\
\hline
\multicolumn{3}{l}{\parbox{8cm}{\textbf{Notes.} 
$(1)$ Torres 2007; $(2)$ H\"unsch \etal\ 1999, Sanz-Forcada \etal\ 2010, Poppenhaeger \etal\ 2010; (3) Demory \etal\ 2007; (4) \emph{Hipparcos} catalogue, Perryman \etal\ 1997.}}
\end{tabular}
\end{table}

\section{Observations and data reduction}
%========================================
\label{sec:obs}

We obtained \hst\ time (GO/DD\#11817) to precisely estimate the \Lya\ emission of the M dwarf GJ~436. The data were recorded on 2010 January 6 with the STIS instrument (Woodgate \etal\ 1997). The observation consists of one 1762-s exposure on the Far Ultraviolet Multi-Anode Microchannel Array detector (FUV-MAMA). The STIS/FUV-MAMA is a solar-blind cesium iodide (CsI) detector with a $25\arcsec\times25\arcsec$ field of view ($\sim 0\farcs025$~pixel$^{-1}$), operating from 1\,150 to 1\,700~\AA. The light was diffracted using the long slit of size $52\arcsec\times0\farcs1$ and the first-order grating G140M. This grating was tilted to ensure a central wavelength of 1\,222~\AA. In this configuration, our data have a spectral coverage of 1\,194--1\,249~\AA\ with a medium resolution of $\sim 10\,000$ ($30$~km~s$^{-1}$) and a throughput of between $\sim 1.2\%$ and $2\%$ (see Proffitt \etal\ 2010). The data were recorded in time-tag mode, and treated with the version 2.27 of \texttt{CALSTIS}, the STIS data pipeline. \texttt{CALSTIS} performs all the standard data reduction tasks (such as bias and dark subtraction, flat fielding, and wavelength and flux calibrations; see Dressel \etal\ 2007 for more details). For spectroscopy, the pipeline end-products for a single exposure are the two-dimensional spectral image and the one-dimensional spectrum.

\begin{figure}
%+++++++++++++
\begin{center}
\resizebox{\columnwidth}{!}{\includegraphics{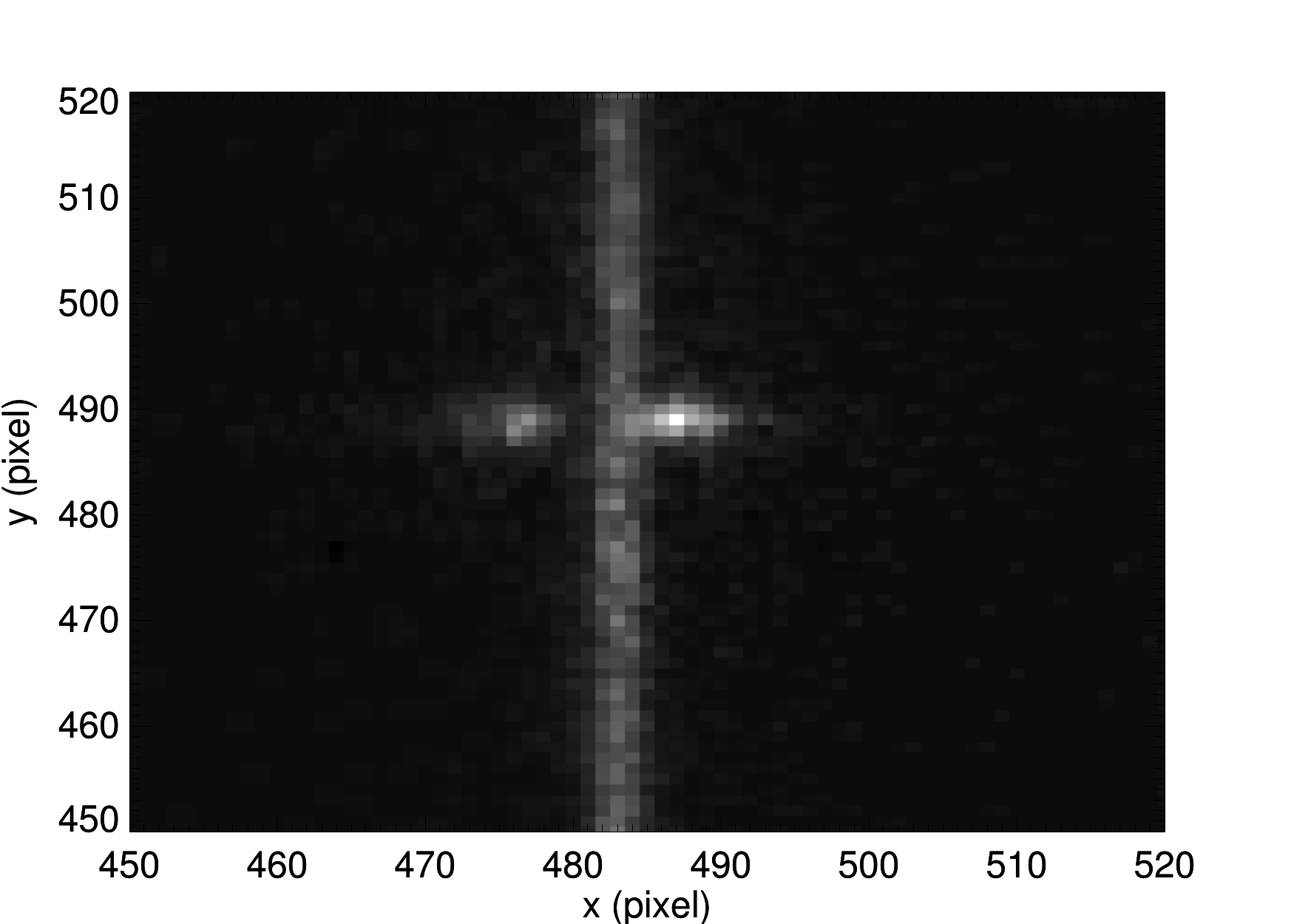}}\caption{\label{fig:x2d} Two-dimensional spectral image of GJ~436. The light is dispersed along the x-axis (roughly). The stellar \Lya\ emission is seen on both sides of the geocoronal emission, which impacts the detector across the whole y-axis.}
\end{center}
\end{figure}

\begin{figure}
%+++++++++++++
\begin{center}
\resizebox{\columnwidth}{!}{\includegraphics{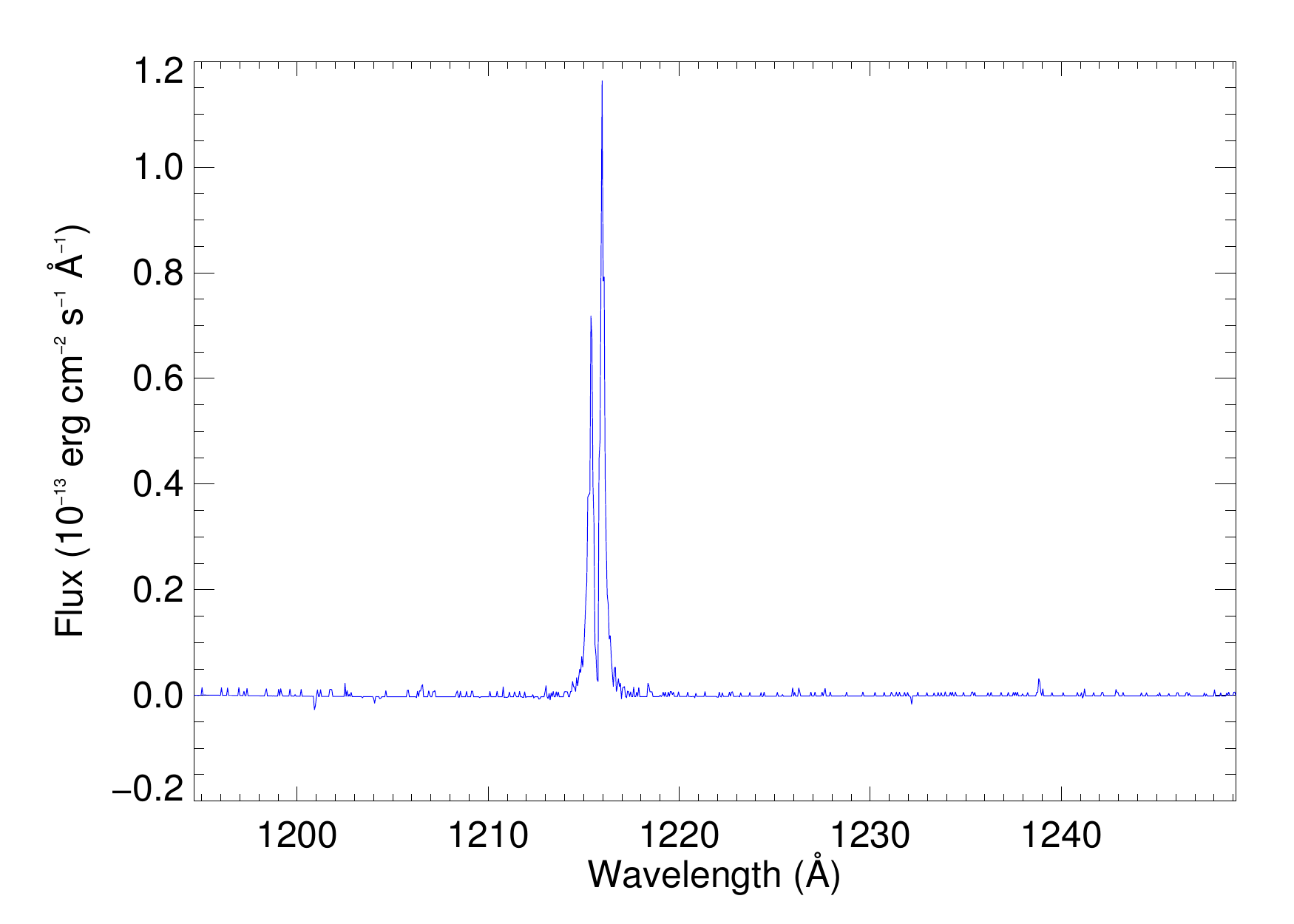}}\caption{\label{fig:x1d} The reduced and calibrated STIS/G140M spectrum of GJ~436 in January 2010.}
\end{center}
\end{figure}

Figure~\ref{fig:x2d} shows the geometrically rectified, wavelength-, and flux-calibrated 2-D spectral image (X2D). The spectral dispersion is along the x-axis. In this image, the stellar \Lya\ emission appears on both sides of the prominent geocoronal emission, which is imaged across the whole detector y-axis. This terrestrial air glow seriously degrades the quality of spectroscopic observations at \Lya; however its rather narrow extent along the dispersion (x-)axis and its large spatial extent (y-axis) enable us to perform  an efficient correction. This task is handled by the pipeline subroutine called \texttt{BACKCORR}, which performs the 1-D spectral background subtraction. The background is determined along a five-pixel-wide region located $\pm300$ pixels away from the centre of the spectral extraction region. The 1-D extracted background is then fitted by a $n$-th degree polynomial, which replaces the background \emph{except} in the regions surrounding the \Lya\ line (Dressel \etal\ 2007). The resulting calibrated 1-D spectrum of GJ~436 (X1D) is shown in Fig.~\ref{fig:x1d}.

\section{Lyman-$\alpha$ emission of GJ~436}
%==========================================
\label{sec:flux}

\subsection{Observed flux at Earth}
%----------------------------------
Figure~\ref{fig:lya} shows a zoomed region of the recorded spectrum around the \Lya\ line. The error bar for each spectral bin is estimated by \texttt{CALSTIS}, which basically propagates the statistical noise $\sqrt(Ng)/g$ in each pixel of the raw data, where $N$ is the number of data events (counts) and $g$ is the gain (equal to 1 photon~count$^{-1}$ for MAMA observations). 

The total flux of the line seen by \hst, integrated from 1\,214.40 to 1\,215.58~\AA\ and from 1\,215.75 to 1\,217.00~\AA\ in order to avoid any residuals from the air glow subtraction at the centre of the line, is $(5.490\pm0.184)\times10^{-14}$~\fluxunitint. We note that, as seen from the Earth, the \Lya\ flux of GJ~436 is higher than that of HD~209458b: the stellar emission from the latter star is also reproduced in Fig.~\ref{fig:lya} for comparison.

The observed shape of the line is a characteristic double peak resulting from the addition of several profiles, as shown in Fig.~\ref{fig:line_profile}. As for the Sun (Woods \etal\ 2005), the core of the intrinsic stellar \Lya\ line is an emission feature originating in the transition region, where the temperature increases quickly between the chromosphere and the corona. The wings of the \Lya\ line originate in a deeper chromospheric region. The emission can be either single-peaked or double-peaked, with a central dip due to the high opacity of the abundant hydrogen atoms. As seen from the Solar System, the central part of the line is strongly absorbed by the ISM from 1\,215.58 to 1\,215.75~\AA . The ISM absorption can indeed be Doppler-shifted relative to the line centre by several kilometers per second. The ISM absorption is the combination of absorption by neutral hydrogen (\ion{H}{i}) and deuterium (\ion{D}{i}). Deuterium produces narrow absorption at 1\,215.3~\AA , which is blue-shifted by $\sim$0.3~\AA\ from the H\,{\sc i} absorption line. The observed \Lya\ profile is therefore caused by the stellar emission line being absorbed by the ISM and subsequently convolved with the instrumental line spread function (Fig.~\ref{fig:line_profile}). 

\subsection{Reconstructed stellar intrinsic emission}
%----------------------------------------------------

To model the hydrogen atom dynamics under the stellar \Lya\ radiation pressure, we need to estimate the stellar emission line as seen by hydrogen atoms escaping from the planet. To calculate the resulting \Lya\ transit light curve, we also need to estimate the absorption by the escaping atoms over the stellar profile as seen from the Earth (Sect.~~\ref{sec:model}). To produce these two line profiles, we fitted the observed \Lya\ profile with a stellar emission line and an ISM absorption model, following the method proposed by Wood \etal\ (2005). In the following, we assume a D/H ratio of $1.5\times10^{-5}$. There is widespread consensus (e.g., Ferlet \etal\ 2000; H\'ebrard \& Moos 2003; Linsky \etal\ 2006) that this value of the D/H ratio is constant within the Local Bubble ($<100$~pc; see Wood \etal\ 2005 and references therein). 

\begin{figure}
%+++++++++++++
\begin{center}
\resizebox{\columnwidth}{!}{\includegraphics{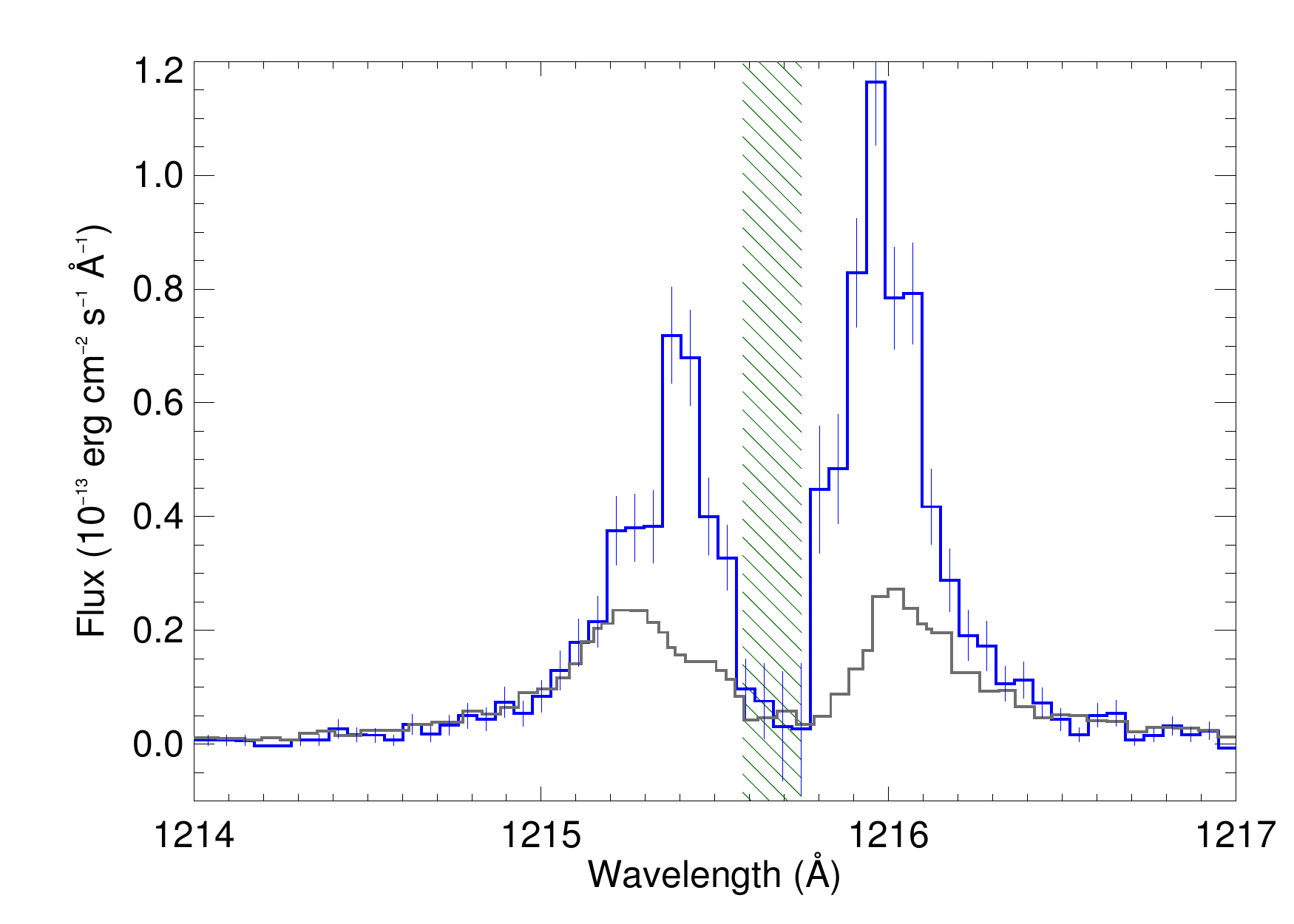}}\caption{Zoom on the \Lya\ line  of GJ~436 measured with STIS/G140M with error bars from the STIS pipeline (blue curve). The STIS spectrum of HD~209458 obtained by Vidal-Madjar \etal\ (2003) is shown for comparison (grey curve). As seen from Earth, the nearby M dwarf clearly shows a more intense stellar emission and a narrower ISM absorption. The location of the air glow (subtracted here) is indicated by green hatches.\label{fig:lya}}
\end{center}
\end{figure}

The ISM column density value is chosen within the range considered by Wood \etal\ (2005), $17.6 < \log N(\textrm{\ion{H}{i}}) < 18.2$. The value is adjusted to $10^{18}$~cm$^{-2}$ using the \ion{D}{i} line at 1\,215.25~\AA. This agrees with the ISM column density of $\log N(\textrm{\ion{H}{i}}) = 17.82$ measured by Wood \etal\ (2005) toward the star HD~4345, in a similar direction ($11.1\degr$ from GJ~436) and at a distance of 21.7~pc. A column density $\log N(\textrm{\ion{H}{i}}) \gg 18$ would be unrealistic for a target as close as GJ~436 (10.2~pc) and yield a stronger \ion{D}{i} absorption than observed. (The presence of this feature can be guessed in the blue part of the observed \ion{H}{i} \Lya\ emission in Fig.~\ref{fig:lya}.) 
In any case, the fit to the observed \Lya\ profile does not depend much on the assumed ISM column density between $N(\textrm{\ion{H}{i}})\sim10^{17}$ and $10^{18}$~cm$^{-2}$ because at these column densities, the \ion{H}{i} absorption line is saturated. The result is more dependent on the assumed profile of the wings extrapolated to the core of the emission line.

As pointed out by Wood \etal\ (2005), the reconstruction of the intrinsic \Lya\ emission of a star can be affected by absorption from both the heliosphere and the astrosphere, which we did not model for GJ~436. Owing to the moderate resolution of the present observations and our lack of knowledge about the ISM absorption, from either \ion{D}{i} or heavier elements (\ion{Fe}{ii} or \ion{Mg}{ii}), this absorption remain unconstrained. However, GJ~436 is located 87.7\degr\ from the upwind direction of the ISM flow seen by the Sun, hence based on Fig.~13 in Wood \etal\ (2005) it is unlikely that heliospheric absorption is detectable. No extra \ion{H}{i} absorption is required besides the ISM component to achieve a good fit of the \Lya\ line. In particular, we did not consider an astrospheric absorption component. Wood \etal\ (2005) found that this absorption was necessary in some case to obtain satisfactory fits to the \Lya\ lines. This does not seem to be the case for GJ~436 given the current data resolution.

\begin{figure}
%+++++++++++++
\begin{center}
\resizebox{\columnwidth}{!}{\includegraphics{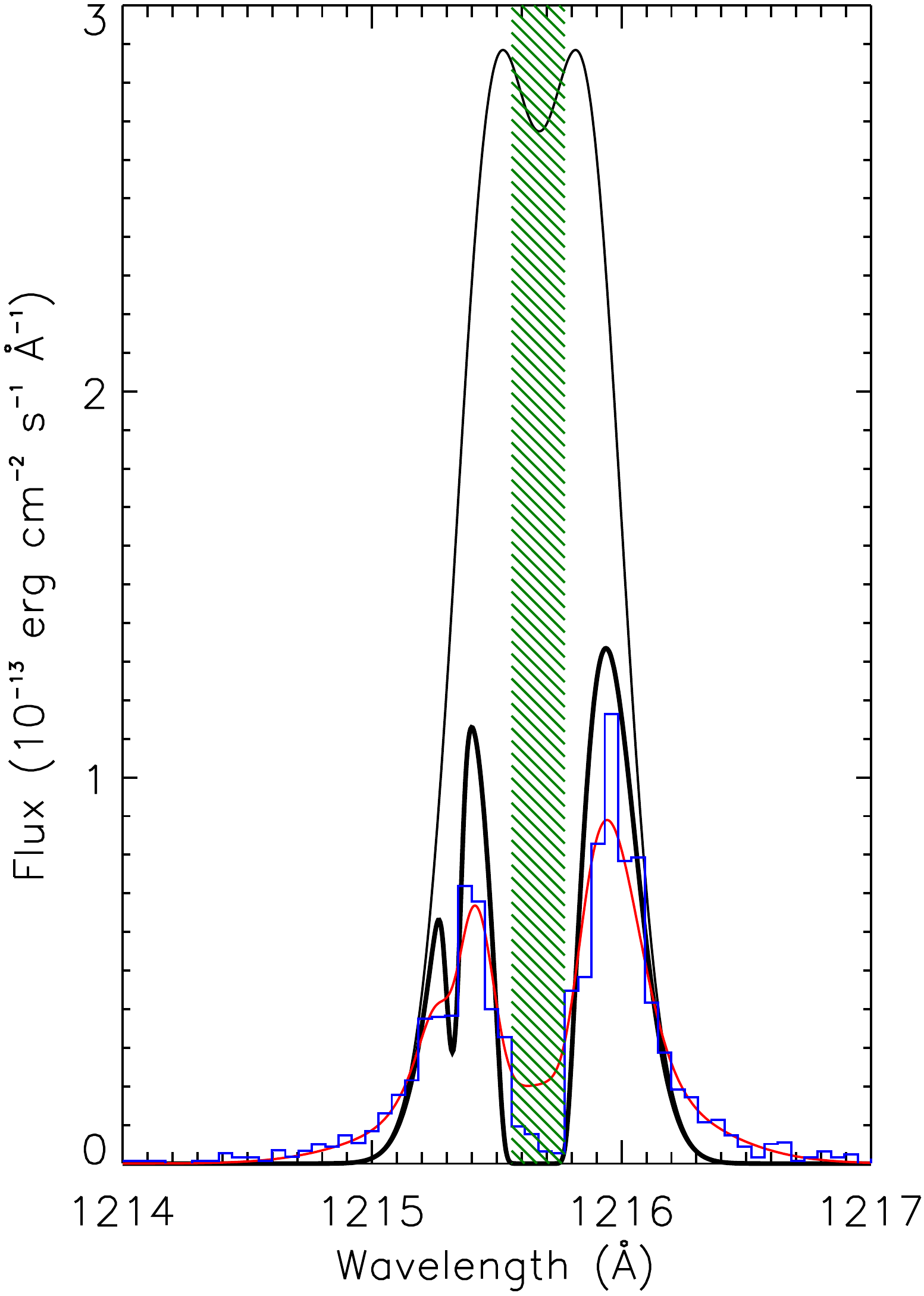}}\caption{Plot of the theoretical profile of GJ~436's \Lya\ line. The black thin line shows the intrinsic stellar emission line as seen by hydrogen atoms escaping the planetary atmosphere. The black thick line shows the resulting profile after absorption by the ISM hydrogen and deuterium. 
The line profile convolved with the \hst\ G140M instrumental line spread function (red line) is compared to the observations (blue histogram), yielding a good fit with a $\chi^2$ of 34.5 for 39 degrees of freedom. The location of the (subtracted) air glow is indicated by green hatches.
\label{fig:line_profile}}
\end{center}
\end{figure}

We performed the model fitting using several intrinsic line profiles, including a single-peak Gaussian and a double-peaked line modelled by a double Gaussian. We found that the double-peaked Gaussian reproduces far more accurately the observed wings in the line profile. The model intrinsic profile is composed of two Gaussians with the same full width at half maximum ($\rm FWHM=\Delta \lambda$) and line centres separated by the same value, $\Delta \lambda$, which is taken as a free parameter. The line total flux is the second free parameter. The ISM and the stellar radial velocities are also free parameters. The model is convolved with the G140M instrumental line spread function, and compared to the observations. We found a satisfactory fit with $\Delta \lambda=0.41$~\AA , an ISM radial velocity of $-4$~km~s$^{-1}$ in the stellar reference frame, and a stellar radial velocity corresponding to a Doppler shift of $0.003$~\AA\ ($\sim0.7$~km~s$^{-1}$). The resulting $\chi^2$ is 34.5 for 39 degrees of freedom (see the data and fit in Fig.~\ref{fig:line_profile}). 

The total flux in the reconstructed \Lya\ line is $(2.7\pm0.7)\times10^{-13 }$~erg~s$^{-1}$~cm$^{-2}$. The error bars were estimated by exploring the parameter space for the fit to the observed \ion{H}{i} line wings, and considering possible single-peaked profiles or the possibility of two narrow peaks with a deep self-absorption in the central part of the emission line. These last two cases produce extreme \Lya\ emission profiles within the ISM absorbed wavelength range; the resulting error bars should be considered as conservative.

\section{Mass-loss rate of GJ~436b}
%==================================
\label{sec:rate}

The mass-loss rate $\dot{m}$ of GJ~436b is a free parameter in our modelling of the evaporation signature described in the next section. Nevertheless, estimating its value is useful to constraining all the parameter space. To achieve this, we use the method presented by Lecavelier des Etangs (2007), which consists of locating the planet in an ``energy diagram''. The energy diagram of exoplanets measures the whole gravitational potential energy of a planet, as a function of high-energy irradiations it receives. This diagram was updated by Davis \& Wheatley (2009), and is being completed with a larger sample of transiting planets (Ehrenreich \& D\'esert 2011), including GJ~436.

\begin{figure}
%+++++++++++++
\begin{center}
\resizebox{\columnwidth}{!}{\includegraphics{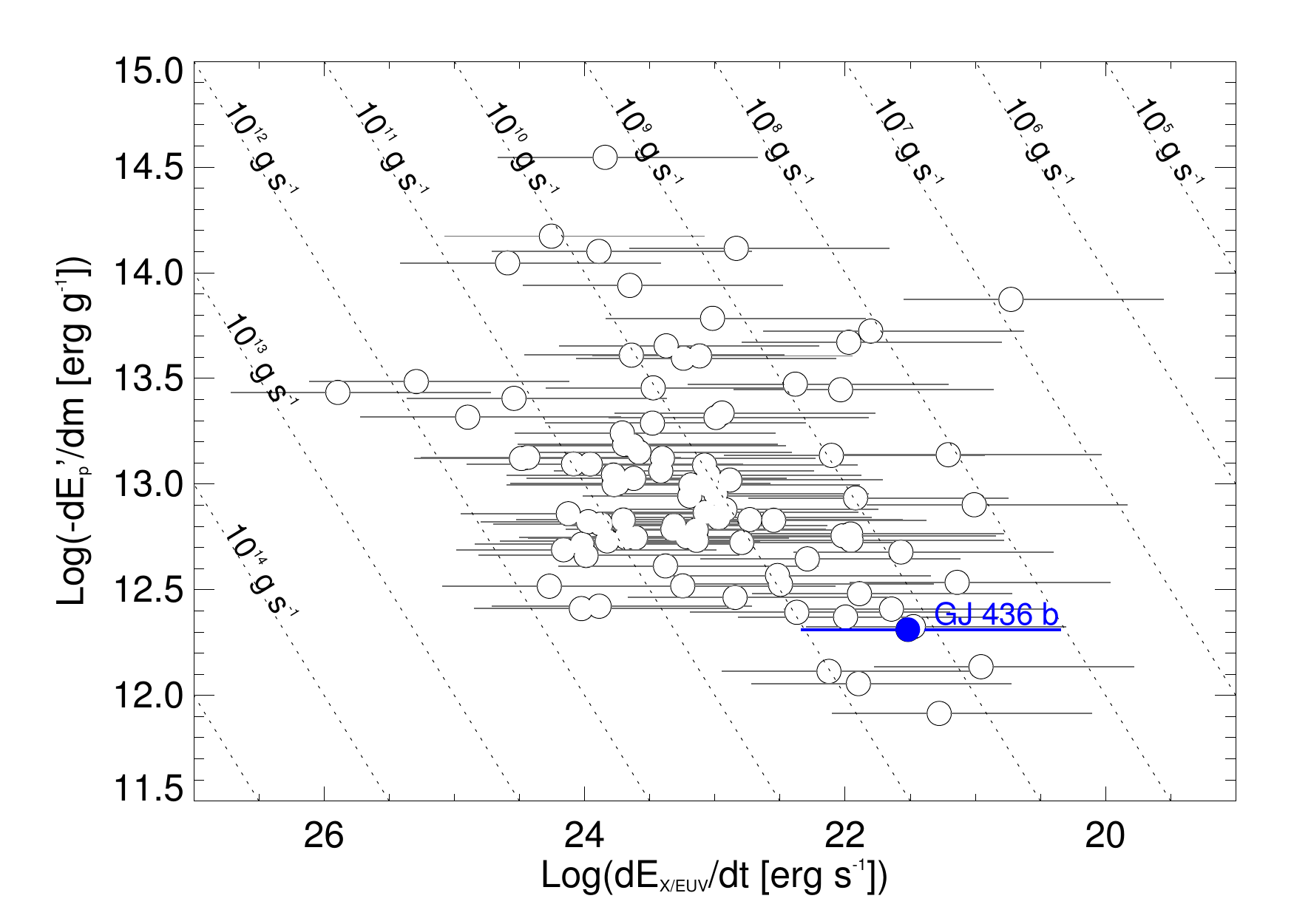}}
\caption{Energy diagram for 100 transiting exoplanets, updated from Lecavelier des Etangs 2007. The energy needed to escape a unit of mass of the planet atmosphere is plotted versus the X/EUV flux. The dotted lines indicate constant mass-loss rates. The blue dot indicates the location of GJ~436b, calculated here assuming an heating efficiency of $\eta=0.15$. The horizontal error bars represent variations in $\eta$ of between 0.01 and 1.} \label{fig:Ediag}
\end{center}
\end{figure}

The resulting diagram is presented in Fig.~\ref{fig:Ediag}. For GJ~436, we calculate, following Lecavelier des Etangs (2007), the potential energy per mass unit of the planet
\begin{equation} 
%///////////////
\label{eq:dEpot}
\frac{\dd E'_p}{\dd m} = -\G M_p / R_p + \delta_\mathrm{tides} = -2.05\times10^{12}~\mathrm{erg~g^{-1}},
\end{equation}
where $\G$ is the gravitational constant and $\delta_\mathrm{tides}$ is the modification of the planet gravitational potential by the stellar tides (see Appendix B of Lecavelier des Etangs 2007). 

The energy received by unit of time by GJ~436b can be expressed as
\begin{equation} 
%///////////////
\label{eq:dEeuv}
\frac{\dd E_\mathrm{X/EUV}}{\dd t} = \frac{\eta}{4} R_p^2 a_p^{-2} L_\mathrm{X/EUV} = 2.19\times10^{22}\eta~\mathrm{erg~s^{-1}}.
\end{equation}
The `X/EUV' in $L_\mathrm{X/EUV}$ means that the luminosity should be integrated from $\sim1$~\AA\ to 912~\AA, following Cecchi-Pestellini \etal\ (2009).\footnote{In this paper, we use the following conventions for naming spectral regions: X-rays (X) in 1--10~\AA, soft X-rays (XUV) in 10--30~\AA, extreme ultraviolet (EUV) in 30--912~\AA, and far ultraviolet (FUV) in 912--1216~\AA.} However, observations do not usually cover a sufficiently wide bandpass to yield a precise value of $L_\mathrm{X/EUV}$. The \emph{Rosat} luminosity (H\"unsch \etal\ 1999) given in Table~\ref{tab:gj436b} is calculated from 0.1 to 2.4~keV, i.e., from 5 to 124~\AA. It is, therefore, a lower limit to the actual X/EUV luminosity.  

The factor $\eta$ is the heating efficiency in the exoplanet thermosphere. Lecavelier des Etangs (2007) considered the extreme case of $\eta = 1$ where all the stellar flux is used to escape the atmosphere. Considering (without quantifying) energetic losses due to thermal emission by atmospheric hydrogen, Tian \etal\ (2005) modelled the atmospheric escape process of the hot Jupiter HD~209458b by assuming that $\eta=0.15$. This value was initially chosen by Watson, Donahue \& Walker (1981) in their pioneering study of Earth's atmospheric escape. In the following, we consider both of these values of $\eta$. 

The resulting mass-loss rate is thus
\begin{equation}
%///////////////
\dot{m} = -\eta \frac{\dd E_\mathrm{X/EUV}}{\dd E'_p}.
\end{equation}
For $\eta = 1$ and $0.15$, we find that $\dot{m} = 1.07\times10^{10}$ and $1.60\times10^9$~g~s$^{-1}$, respectively. These values are close to the canonical escape rate predicted by several evaporation models (e.g., Yelle 2004, 2006; Lecavelier des EtangsÊ\etal\ 2004; Garc\'\i a-Mu\~noz 2007; Murray-Clay, Chiang \& Murray 2009).

\section{Observable signature of the evaporating atmosphere}
%===========================================================
\label{sec:model}

\begin{figure}
%+++++++++++++
\begin{center}
\resizebox{\columnwidth}{!}{\includegraphics[angle=90]{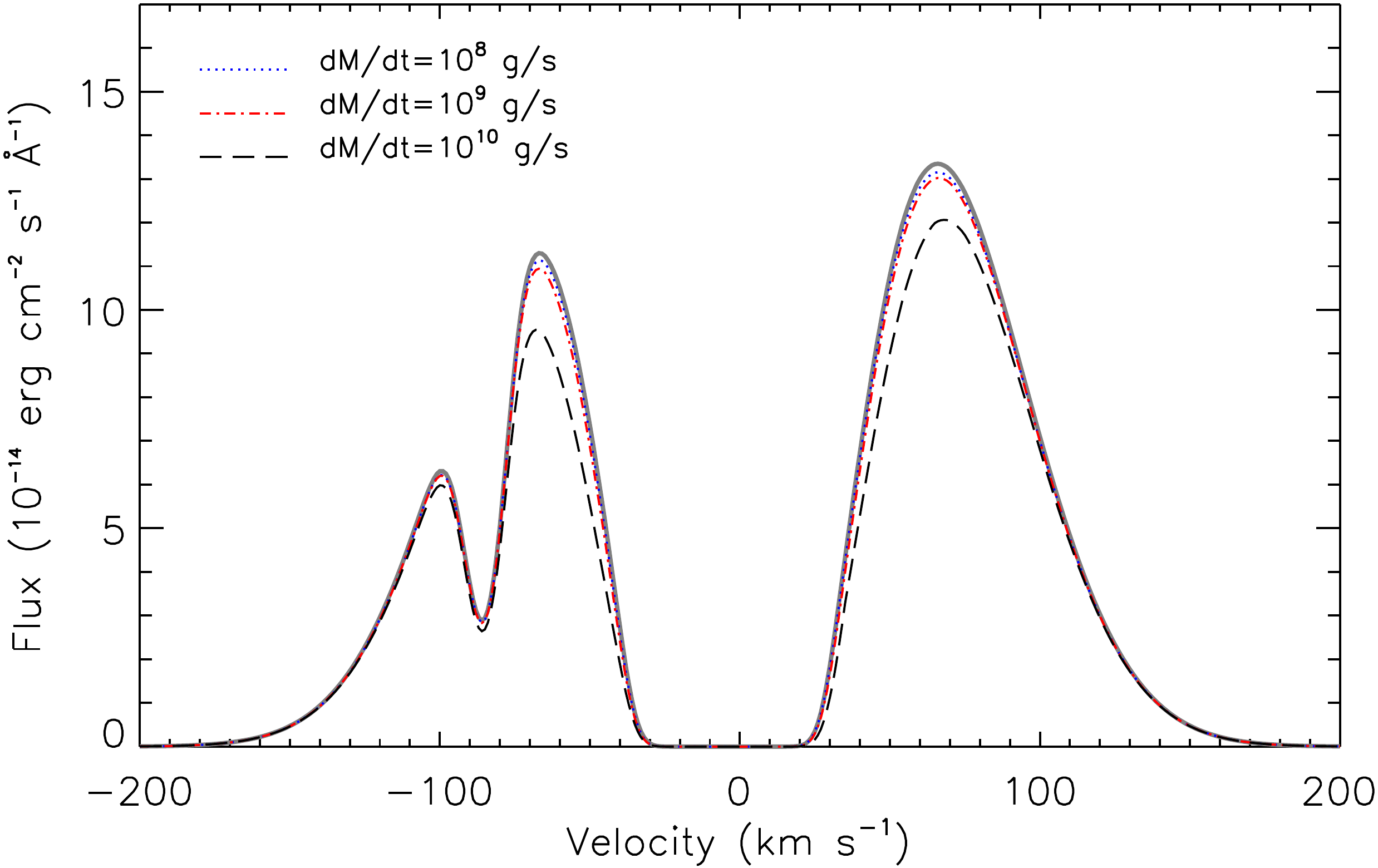}} 
\caption{The GJ~436 \Lya\ spectrum during the transit of GJ~436b. 
The star spectrum before or after the transit is shown by a thick grey line. 
The spectrum resulting from the absorption by the planetary exophere  during the transit
is shown for various escape rates. The blue dotted, the red dot-dashed, and black long-dashed lines show the resulting spectrum when the escape rate is 10$^{8}$, 10$^{9}$, and 10$^{10}$~g~s$^{-1}$, respectively.
With this last escape rate, the blue side of the line is absorbed by about 17\%,
resulting in a decrease of about 11\% in the total Lyman-$\alpha$ flux.
\label{fig:abs_profile}}
\end{center}
\end{figure}

\begin{figure}
%+++++++++++++
\begin{center}
\resizebox{\columnwidth}{!}{\includegraphics[angle=90]{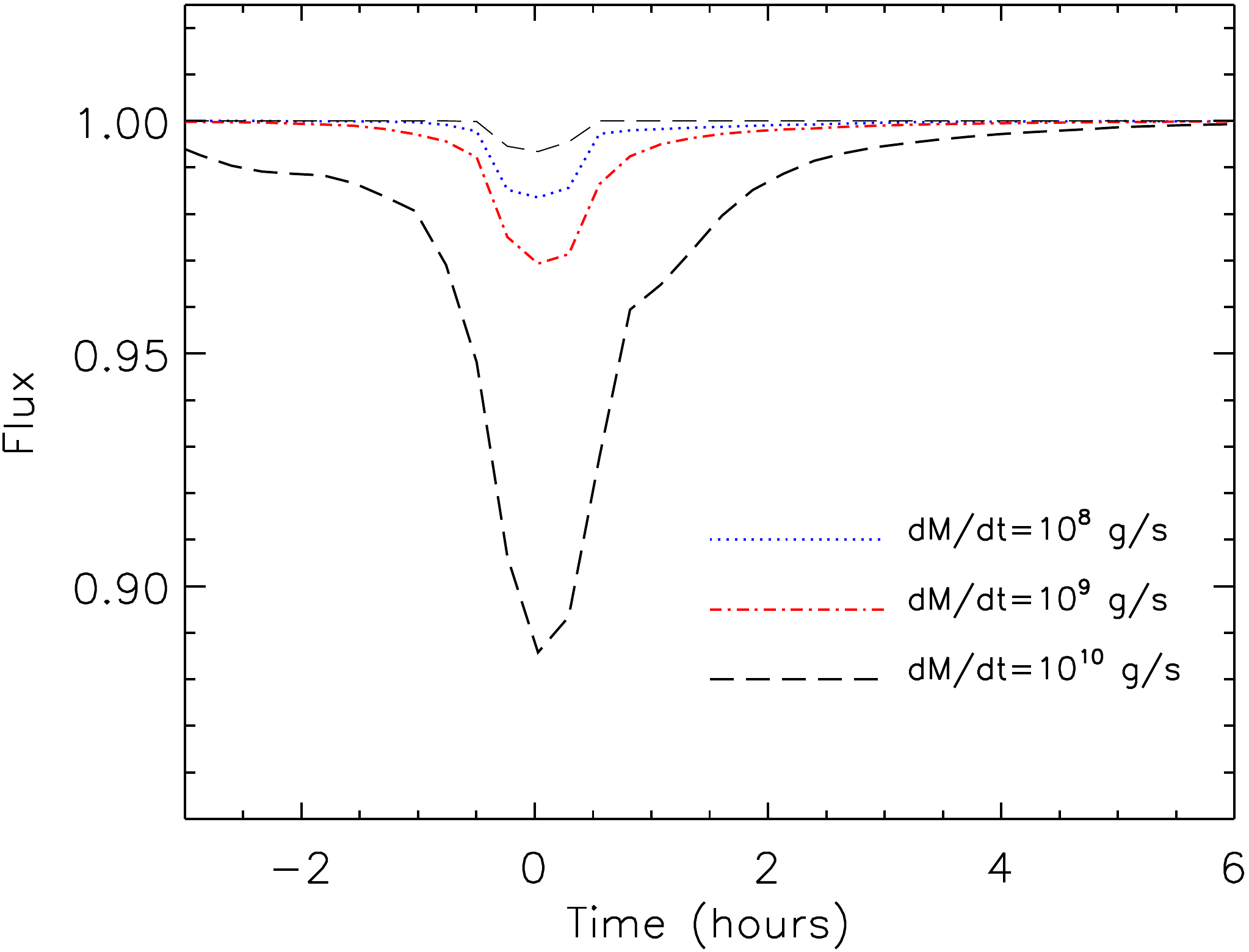}}
\caption{Plot of the theoretical light curve of the total \Lya\ flux 
for various escape rates $\dot{m}$: 10$^{8}$~g~s$^{-1}$ (blue dotted line), 10$^{9}$~g~s$^{-1}$ (red dot-dashed line), and 10$^{10}$~g~s$^{-1}$ (black dashed line). The thin long-dashed line shows the light curve for the planet without additional absorption by the planetary atmosphere.
\label{fig:F_vs_t}}
\end{center}
\end{figure}

To calculate a theoretical H\,{\sc i} transit light curve for various atmospheric escape rates, we modelled the atmospheric gas escaping GJ~436b with a numerical simulation including the dynamics of the hydrogen atoms. In this $N$-body simulation, hydrogen atoms are released from GJ~436b's upper atmosphere as particles with a random initial velocity corresponding to a 10\,000-K exobase. This temperature corresponds to the temperature expected in the upper atmospheres of hot jupiters (e.g., Lecavelier des Etangs \etal\ 2004; Stone \& Proga 2009) and was inferred from observations (Ballester, Sing \& Herbert 2007; Vidal-Madjar \etal\ 2011). In any case, our results do not depend on this assumption because atoms are rapidly accelerated by the radiation pressure to velocities several times higher than their initial velocities. In this model, the gravity of both the star and the planet are taken into account. The radiation pressure from the stellar \Lya\ emission line is calculated as a function of the radial velocity of the atoms. The self-extinction within the cloud is also taken into account. The hydrogen atoms are supposed to be ionized by the stellar EUV flux ($<912$~\AA). In the simulation, their lifetime is calculated as a function of the ionizing EUV flux which is taken to be equal to the solar value. The only free parameter is the atomic hydrogen escape rate, expressed in grams per second. The dynamical model provides a steady state distribution of positions and velocities of escaping hydrogen atoms in the cloud surrounding GJ~436b. From this information, we calculated the corresponding absorption over the stellar line (Fig.~\ref{fig:abs_profile}) and the corresponding transit light curve of the total \Lya\ line (Fig.~\ref{fig:F_vs_t}). We found that escape rates of $\sim10^9$ to $10^{10}$~g\,s$^{-1}$ can produce absorptions of between $\sim3\%$ and $11\%$  in the \Lya\ transit light curve when integrating over the whole line. Absorption depths this large could be promptly detected with \hst\ \Lya\ observations. 

This contrasts with the atmospheric signatures expected in the infrared.
The atmospheric scale height of this planet is $H=\kB T_B / (\mu g_p) \approx 232$~km, where $\mu = 2$~g~mol$^{-1}$ is the molar mass of a H$_2$ atmosphere and $g_p$ is the surface gravity of the planet. We can expect atmospheric transit signatures at the level of $\sim 2 (\Delta F/F) (H / R_p)$ (Winn 2010), where $\Delta F/F \approx (R_p/R_\star)^2 \approx 0.7\%$ is the transit occultation. This yields a typical extra absorption value of $\sim10^{-4}$ for one atmospheric scale height. This value is consistent with the upper limits on the atmospheric absorption of GJ~436b set by Pont \etal\ (2009), who have covered two near-infrared transits of the planet with the NICMOS camera on \hst\ using a grism to sample the 1.1--1.9~\micron\ band. These authors report observing a flat transmission spectrum at the level of $\sim 10^{-4}$, and in particular no significant signal in the 1.4-\micron\ water band (see also Gibson, Pont \& Aigrain 2010). 

Using eclipses of the planet by the star observed by \emph{Spitzer}, Stevenson \etal\ (2010) obtained planet-to-star flux ratios that show marginal variations between 3.6 and 24~$\mu$m (see also Beaulieu \etal\ 2011). The interpretation of these broad-band spectrophotometry data is difficult and relies on statistical approaches (Madhusudhan \& Seager 2011). Shabram \etal\ (2011) modelled the infrared transmission spectrum of GJ~436b. These authors conclude that the detailed infrared characterization of this planet may be possible using the \emph{James Webb Space Telescope}. Before its launch, the use of UV facilities should be considered as a powerful way of characterizing the upper atmospheres of low-mass exoplanets. 

\section{\Lya\ emission of M dwarfs and stellar activity}
%=========================================================
\label{sec:Mdwarfs}

Stellar \Lya\ emission lines are important spectral features in the context of exoplanet stellar environment and stellar physics. These emission lines are the main contributors to the flux of low-mass stars at short wavelengths, from X to FUV. For instance, the \Lya\ emission of the Sun contributes to more than 50\% of its flux below 1\,216~\AA\ (Woods \etal\ 1998). The \Lya\ line is used as a proxy for determining the temperature and pressure profiles of upper stellar atmospheres. 

The \Lya\ emission originate in the transition regions between the chromospheres and coron\ae, while other commonly used activity diagnostics are based on indicators originating from the coron\ae\ (X-ray flux) or the chromospheres (\ion{Mg}{ii} $h$ and k lines). In this context, stellar \Lya\ emission provides an almost unique access to the structure of the transition region, where the temperature profile exhibits a deep minimum. Houdebine \& Doyle (1994) modelled the \Lya\ emission of M dwarfs. According to their work, the transition region of these stars is thinner than that of solar-like stars.

Stellar \Lya\ measurements are scarce in the literature. This is mainly because it is impossible to observe this emission from the ground and that it is difficult to correct measurements for the effects of the ISM absorption, both astrospheric and heliospheric absorption, and geocoronal emission. Furthermore, correlations between \Lya\ emission and other proxies of stellar activity can be misleading because they do not originate in the same region. This ensures that an accurate estimate of the \Lya\ brightness from an activity relation is rather hazardous. In addition, the \Lya\ line is extremely variable with time. For the Sun, the integrated \Lya\ line flux may change by 37\% during one rotation and up to 50\% over a couple of years (Vidal-Madjar 1975). During the magnetic cycle, the extreme (single day) values can vary by more than a factor of two (Woods \etal\ 2000).

Very few M dwarfs have \Lya\ measurements. Landsman \& Simon (1993) used the \emph{International Ultraviolet Explorer} (\iue) to report \Lya\ fluxes corrected from the ISM absorption for 12 M dwarfs. Wood \etal\ (2005) obtained resolved \Lya\ measurements for four M dwarfs with \hst/STIS. Three objects are common to both samples, leading to a total of 13 measurements. This sample consists mostly of active stars. In particular, all the \hst\ measurements of Wood \etal\ (2005) were obtained for M dwarfs with magnetic activity levels well above that of GJ~436. Therefore, our measurement of GJ~436's \Lya\ flux is invaluable for estimating the properties of the transition region in quiet M dwarfs.

\begin{figure}
%+++++++++++++
\begin{center}
\resizebox{\columnwidth}{!}{\includegraphics{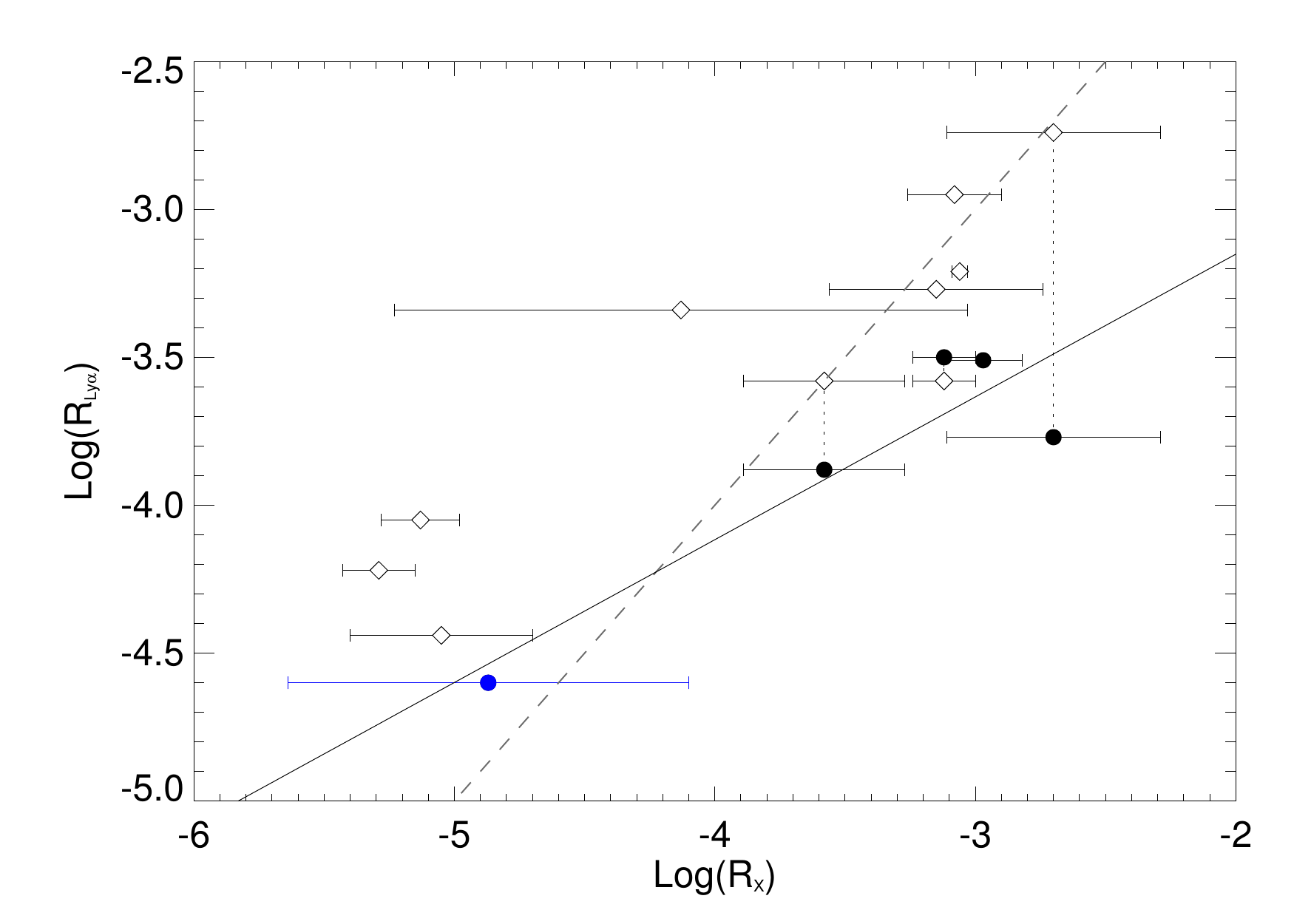}} 
\caption{Normalized \Lya\ flux $R_\mathrm{Ly\alpha}=f_\mathrm{Ly\alpha}/f_\mathrm{bol}$ as a function of normalized X-ray flux $R_{X}=f_{X}/f_\mathrm{bol}$ for all the \iue\ ($\diamond$) and \hst\ ($\bullet$) \Lya\ flux measurements of M dwarfs in the literature. Our measurement for GJ~436 data is shown in blue. The dotted lines link \hst\ and \iue\ mesasurements for objects that have been measured with both instruments. The plain line show the best-fit linear regression to \hst\ measurements, $\log(R_\mathrm{Ly\alpha}) \approx 0.5 \log(R_{X}) - 2.2$. The grey dashed line indicates the $\log R_\mathrm{Ly\alpha} = \log R_X$ relation. The horizontal error bars in the X-ray flux are the dispersions reported in Table~\ref{tab:rlya_rx}. \label{fig:rlya_rx}}
\end{center}
\end{figure}

The measurements found in the literature are listed in Table~\ref{tab:rlya_rx}. In Fig.~\ref{fig:rlya_rx}, we have plotted the normalized \Lya\ flux, $R_\mathrm{Ly\alpha} = f_\mathrm{Ly\alpha} / f_\mathrm{bol}$ as a function of the normalized X-ray flux $R_X = f_X / f_\mathrm{bol}$ for the \emph{IUE} and \hst\ measurements from Table~\ref{tab:rlya_rx}. We have not included the available data for \object{YY~Gemini} and \object{DT~Virginis} because these stars are close binaries for which a definitive identification of each component's flux is difficult. The values are listed in Table~\ref{tab:rlya_rx}. The observed X-ray fluxes $f_X$ are extracted from the \emph{Rosat} all-sky survey (RASS) in the Nexxus~2 database (Schmitt \& Liefke 2004). The bolometric fluxes $f_\mathrm{bol}$ are obtained by combining the $IK$ photometry of Leggett (1992) with the bolometric correction $BC_K$ versus $I-K$ relation of Leggett \etal\ (2000).

As can be seen in Fig.~\ref{fig:rlya_rx}, the \Lya\ fluxes extracted from \hst\ are significantly lower --\,for a given $R_X$\,-- than those from \emph{IUE}. Wood \etal\ (2005) note that the \hst\ and \emph{IUE} data are coherent for the majority of the objects (besides M dwarfs), except for the fainter ones for which the \emph{IUE} fluxes are systematically higher than the \hst\ ones. They attribute this to a bias in the corrections for the ISM absorption and geocoronal emission, which are particularly difficult at the lower resolution of \emph{IUE}.

In the case of faint M dwarfs, we prefer to consider the clearly resolved \hst\ measurements, which were obtained mainly for magnetically active stars. The values of $R_\mathrm{Ly\alpha}$ seem to be correlated to $R_X$ for M dwarfs, with $\log R_\mathrm{Ly\alpha} \approx 0.5\log R_X - 2.2$, although additional measurements are necessary to confirm this trend. The trend implies that $R_\mathrm{Lya}/R_X > 1$ for quiet M dwarfs, in contrast to the results for more active ones. Figure~\ref{fig:rlya_rx} can thus be used --\,with caution\,-- to estimate $R_\mathrm{Ly\alpha}$ for all M dwarfs that have values of $R_X$ between $-5$ and $-3$, i.e., covering the domain of magnetic activity between 20\% of that of the quietest M dwarfs in the solar neighbourhood to the very active stars close to the X-ray emission saturation level (see for instance Fig.~5 in Delfosse \etal\ 1998).

Meanwhile, we stress that the X-ray and \Lya\ fluxes have been measured here at different times and, therefore, for different magnetic states of the stars, which can add dispersion to any relationship. For GJ~436, the measured $\log L_X$ ranges from 25.96, as measured with \emph{XMM-Newton} by Sanz-Forcada \etal\ (2010), to 27.15, as found in the \emph{Rosat}/PSPC archives by Poppenhaeger \etal\ (2010). The value we have used throughout this work comes from the \emph{Rosat} all-sky survey catalogue of the nearby stars (H\"unsch \etal\ 1999) and is in-between these extreme values. The dispersion (standard deviation) in these three values is $\sigma_{L_X} = 0.77$.

We have retrieved all available X-ray measurements for the stars in Table~\ref{tab:rlya_rx} from the Nexxus~2 database and calculated their $\sigma_{L_X}$. However, these calculated dispersions only provide a rough idea of the real intrinsic X-ray variabilities of the listed M dwarfs, because the number of measurements (also reported in Table~\ref{tab:rlya_rx}) is not equivalent for all stars. In addition, the measurements are derived using different instruments, mainly \emph{Rosat} or \emph{XMM-Newton} (or both). Differences between surveys, e.g.\ in the bandpasses, are not taken into account, thus adding to the dispersion. Finally, the occurrence of a flare during a given observation cannot be excluded, leading to a non-gaussian distribution of the measurements. Therefore, these values should only be regarded as indicative.

\begin{table}
%************
\begin{center}
\caption{Normalized \Lya\ and X-ray fluxes of M dwarfs.\label{tab:rlya_rx}
}
\begin{tabular}{llllll} 
\hline \hline
Star & Sp.~T. & $R_\mathrm{Ly\alpha}$ & ${R_{X}}^*$ & $\sigma_{L_X}$ (dex) & $N_X$ \\ 
\hline
\multicolumn{5}{c}{\Lya\ from \hst} \\
\hline
\object{Proxima~Cen} & M6{\sc v}   & $-3.88$ &  $-3.58$ & $0.31^\dagger$            & 10 \\
\object{AD~Leo}      & M4.5{\sc v} & $-3.50$ &  $-3.12$ & $0.12^\dagger$            & 6  \\
\object{EV~Lac}      & M3.5{\sc v} & $-3.77$ &  $-2.70$ & $0.41^\dagger$            & 8  \\
\object{AU~Mic}      & M1{\sc v}   & $-3.51$ &  $-2.97$ & $0.15^\dagger$            & 3  \\
GJ~436               & M2.5{\sc v} & $-4.60$ &  $-4.87$ & $0.77^\S$ & 3 \\ 
\hline
\multicolumn{5}{c}{\Lya\ from \iue} \\
\hline
\object{AX~Mic}      & M0{\sc v}   & $-4.22$ &  $-5.29$ & $0.14^\dagger$            & 3  \\
\object{DT~Vir}      & M2{\sc v}   & $-3.27$ &  $-3.15$ & $0.41^{\dagger,\ddagger}$ & 5  \\
\object{GJ~887}      & M2{\sc v}   & $-4.05$ &  $-5.13$ & $0.15^\dagger$            & 2  \\
\object{GJ~411}      & M2{\sc v}   & $-4.44$ &  $-5.05$ & $0.35^\ddagger$     & 4  \\
AD~Leo               & M4.5{\sc v} & $-3.58$ &  $-3.12$ & $0.12^\dagger$            & 6  \\
EV~Lac               & M3.5{\sc v} & $-2.74$ &  $-2.70$ & $0.41^\dagger$            & 8  \\
\object{YZ~CMi}      & M4.5{\sc v} & $-3.21$ &  $-3.06$ & $0.03^{\dagger,\ddagger}$ & 2  \\
\object{V1216~Sgr}   & M3{\sc v}   & $-3.34$ &  $-4.13$ & $1.10^{\dagger,\ddagger}$ & 3  \\
\object{V998~Ori}    & M3.5{\sc v} & $-2.95$ &  $-3.08$ & $0.18^\dagger$            & 2  \\
Proxima~Cen          & M6{\sc v}   & $-3.58$ &  $-3.58$ & $0.31^\dagger$            & 10 \\ 
\hline
\multicolumn{6}{l}{\parbox{8cm}{
\textbf{Notes.} ${(*)}$ Normalized X luminosities are \emph{Rosat} values extracted from the Nexxus~2 data base. The dispersion $\sigma_{L_X}$ is the standard deviation in the $N_X$ measurements found in the Nexxus~2 database. These measurements are derived using $(\dagger)$ \emph{Rosat}, $(\ddagger)$ \emph{XMM-Newton}, or $(\dagger,\ddagger)$ both. $(\S)$ The dispersion in the case of GJ~436 is obtained as the standard deviation in the three $L_X$ values listed in the text.}}
\end{tabular}
\end{center}
\end{table}

\section{Conclusion}
%===================
\label{sec:conclu}

We have obtained the first UV glance at a star hosting a transiting hot neptune. The \hst\ data unambiguously show that the early and quiet M dwarf GJ~436 has bright \Lya\ emission. With an escape rate of around $\sim10^{10}$~g~s$^{-1}$, the hydrogen upper atmosphere of GJ~436b should cause large absorption ($\sim11\%$) in the \Lya\ transit light curve. These figures will allow the existence of such an extended upper atmosphere to be tested with a dedicated \hst\ program. These results also suggest that other quiet M dwarfs are brighter at \Lya\ than in X-rays. This prediction opens new perspectives for the atmospheric characterization of other hot-neptune and even super-earth atmospheres. Interestingly, the nature of GJ~1214b, a super-earth, or `small-neptune' also orbiting a quiet M dwarf (Charbonneau \etal\ 2009), could be resolved with this technique, provided the star is bright enough at \Lya. 

\begin{acknowledgements}
We are particularly grateful to J.-M. D\'esert and A.~Vidal-Madjar for their precious help with the preparation of the \hst\ observations and comments on the manuscript. We thank the anonymous referee for a helpful review. We would also like to thank  X.~Bonfils, R.~Ferlet, T.~Forveille, G.~H\'ebrard, A.-M. Lagrange, N.~Meunier, and D.~K.~Sing for stimulating discussions. We also thank M.~Mountain for awarding us \hst\ DD time that made this work possible. D.E. is supported by the Centre National d'\'Etudes Spatiales (CNES).
\end{acknowledgements}

\end{document}